\documentclass[a4paper]{revtex4}
\usepackage{graphicx}
\usepackage{fancyhdr}
\usepackage{amsmath}
\newcommand{\eqn}[1]{&\hspace{-0.6em}#1\hspace{-0.6em}&}

\newcommand{\GeV}{\mbox{~GeV}}

\usepackage{color}

\begin{document}

%Title of paper
\title{Higgs Inflation and Scalar Dark Matter
with Right-handed Neutrinos
}

% Repeat the \author .. \affiliation  etc. as needed
%
% \affiliation command applies to all authors since the last
% \affiliation command. The \affiliation command should follow the
% other information

\author{Hiroyuki Ishida}
\affiliation{  Graduate School of Science and Engineering, Shimane University, \\Matsue, 690-8504 Japan
}

\begin{abstract}
We investigate the possibility of Higgs inflation with an extended standard model by right-handed neutrinos and scalar dark matter. 
We find that the masses of the dark matter and one of the right-handed neutrinos should stand around TeV and $10^{-14}$ GeV scale respectively even if the magnitude of tensor-to-scalar ratio indicates sufficiently small value. 
\end{abstract}

%\maketitle must follow title, authors, abstract
\maketitle

\thispagestyle{fancy}

% body of paper here - Use proper section commands
% References should be done using the \cite, \ref, and \label commands
% Put \label in argument of \section for cross-referencing
%\section{\label{}}

%%%%%%%%%%%%%%%%%%%%%%%%%%%%%%%%%%
\section{Introduction}
The discovery of the Higgs boson with wonderful precision of its mass is one of the biggest discoveries for the particle physics in this decade. 
The standard model (SM) of the particle physics has been finally completed by this.
However, we should wonder that whether the Higgs boson is the last ingredient of the particle physics. 
The answer is obviously no. 
There are many problems to answer from the viewpoint of observations. 
One of them is the origin of inflation. 

Higgs inflation is one of attractive scenarios for the answer. 
If this scenario is true, the particle contents might become minimal. 
However, when we assume that a model explain the inflation within SM particles, 
it is well known that the top mass should be slightly smaller than the observational one and non-minimal coupling to the gravity should be quite large \cite{Hamada:2013mya}.
On the other hand, the Ref.~\cite{Haba:2014zda} points out that an extended model by scalar dark matter (DM) and right-handed neutrinos can explain required values for the e-foldings, scalar spectral index, and sizable tensor-to-scalar ratio with the Higgs and top masses which are consistent with the experimental values.

In this work, we reanalyze the possibility of the Higgs inflation in the extended SM by the right-handed neutrinos and the scalar DM. by solving the renormalization group equations (RGEs) at 2-loop level for the quantum corrections to the coupling constants ($\beta$-functions).

%%%%%%%%%%%%%%%%%%%%%%%%%%%%%%%%%%
\section{Model}
In this work, we consider the extended SM by a real singlet scalar 
and the right-handed neutrinos.
The Lagrangians can be written as
\begin{eqnarray}
\mathcal{L} \eqn{=} 
\mathcal{L}_{\rm SM} + \mathcal{L}_S + \mathcal{L}_N\,,\\
\mathcal{L}_{\rm SM} \eqn{\supset} -\lambda \left( \left|H\right|^2 - \frac{v_{\rm EW}^2}{2} \right)^2\,,\label{Eq:Higgs}\\
\mathcal{L}_S \eqn{=} -\frac{m_S^2}{2} S^2 - \frac{k}{2} \left|H\right|^2 S^2 
- \frac{\lambda_S}{4!} S^4 + \left( {\rm kinetic~ terms} \right)\,,\\
\mathcal{L}_N \eqn{=} - \left( \frac{M_R}{2} \bar{N^c} N + y_N \bar{L} H N + {\rm c.c.} \right)
+ \left( {\rm kinetic~ terms} \right)\,,
\end{eqnarray}
where $\mathcal{L}_{\rm SM}$ is the SM Lagrangian in which the Higgs potential is contained as shown in Eq. (\ref{Eq:Higgs}).
$H$ is the Higgs doublet field and $v_{\rm EW}$ is its vacuum expectation value.
$L$, $S$ and $N$ are the lepton doublet, a SM gauge singlet real scalar and the right-handed neutrino fields, respectively.
The coupling constants $k$, $\lambda_S$ and $y_N$ are 
the portal coupling of $H$ and $S$, the quartic self-coupling of $S$, 
and the neutrino Yukawa coupling constant.
$M_R$ denotes the right-handed neutrino mass.
We assume that the singlet scalar has odd parity under an additional $Z_2$ symmetry.
Hence, we can have a candidate for DM 
when the mass and the portal coupling $k$ of $S$ are taken to be appropriate values.
The observed tiny neutrino masses can be obtained by the conventional type-I seesaw mechanism. 
The heaviest right-handed neutrino with the mass $M_R$ generates one active neutrino mass. 
Others can be obtained by lighter right-handed neutrinos with the smaller Yukawa coupling. 
When the neutrino Yukawa couplings are smaller than the bottom Yukawa coupling, 
the contributions from the neutrino Yukawa couplings to the $\beta$-functions are negligible.
In this work, we consider a case that the neutrino Yukawa coupling of only one generation of 
the right-handed neutrino is effective to the $\beta$-functions. 
We have fixed the active neutrino mass as $0.1$ eV here and hereafter. 

The RGEs are solved from Z boson mass scale, $M_Z$, to the Planck scale, $M_{\rm pl}=2.435 \times 10^{18} {\rm GeV}$.
In this analysis, we divide the energy region into three parts in this region.
The each part can be considered as $M_Z \leq \mu < m_S$, $m_S \leq \mu < M_R$, 
and $M_R \leq \mu \leq M_{\rm pl}$.
The behavior of $\lambda(\mu)$ can be roughly supposed as follows:
At first, the evolution of $k$ becomes small as $k(M_Z)$ becomes small, 
because the $\beta$-function for $k$ is proportional to $k$ itself.
In this case, the evolution of $\lambda (\mu)$ is really close to the SM one.
Second, 
the contribution from the additional scalar pushes up the evolution of $\lambda(\mu)$.
Thus, we can make $\lambda(\mu)$ to be positive up to $M_{\rm pl}$ 
with sufficient magnitude of the additional scalar contribution.
On the other hand, 
the contribution from the right-haded neutrinos pulls down $\lambda(\mu)$ 
in the region of $M_R \leq \mu \leq M_{\rm pl}$. 
\begin{figure}[t]
\begin{center}
\includegraphics[width=6cm,clip]{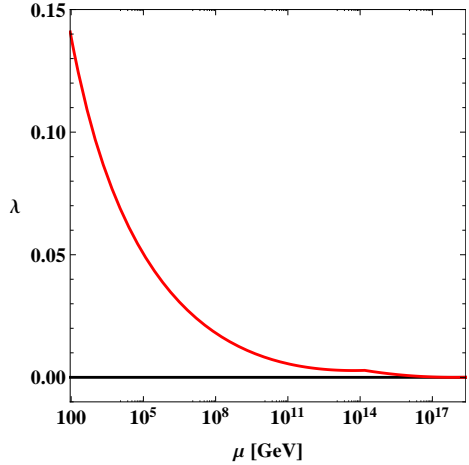}
\caption{A typical behavior of Higgs quartic coupling with central values of Higgs and top masses.}\label{Fig:running}
\end{center}
\end{figure}
A behavior of the running of the Higgs quartic coupling is shown in Fig.\ref{Fig:running}.
As a result, we can have a suitable Higgs potential with a plateau for the inflation 
around $\mathcal{O} (10^{18}) \GeV$~\cite{Haba:2014zda}.
These are the important features for the Higgs inflation with the real scalar field 
and the right-handed neutrinos.\footnote{We assume that $\lambda_S (M_Z)=0.1$ just as a sample point.}
In our estimation, the free parameters are the values of right-handed neutrino and DM masses 
and non-minimal coupling $\xi$.

\section{Higgs inflation with extra singlets}
At first, we briefly review the ordinary Higgs inflation \cite{Bezrukov:2007ep} 
with the action in the so-called Jordan frame, 
\begin{eqnarray}
S_J \eqn{\supset} 
\int d^4 x \sqrt{-g} \left( - \frac{M_{\rm pl}^2 +\xi h^2}{2} R + \mathcal{L}_{\rm SM} \right)\,,
\end{eqnarray}
where $\xi$ is the non-minimal coupling of the Higgs to the Ricci scalar $R$, 
$H=(0\,,h)^T / \sqrt{2}$ is given in the unitary gauge, 
and $\mathcal{L}_{\rm SM}$ includes the Higgs potential of Eq. (\ref{Eq:Higgs}).
With the conformal transformation ($\hat{g}_{\mu \nu} \equiv \Omega^2 g_{\mu \nu}$ 
with $\Omega^2 \equiv 1+\xi h^2/M_{\rm pl}^2$) which denotes the transformation 
from the Jordan frame to the Einstein one, 
one can rewrite the action as 
\begin{eqnarray}
S_E \eqn{\supset} 
\int d^4 x \sqrt{-\hat{g}} \left( - \frac{M_{\rm pl}^2}{2} \hat{R} 
+\frac{\partial_\mu \chi \partial^\mu \chi}{2} 
- \frac{\lambda}{4 \Omega (\chi)^4} \left( h (\chi)^2 - v_{\rm EW}^2 \right)^2 \right)\,,
\end{eqnarray}
where $\hat{R}$ is the Ricci scalar in the Einstein frame given by $\hat{g}_{\mu \nu}$, 
and $\chi$ is a canonically renormalized field as
\begin{eqnarray}
\frac{d\chi}{dh} \eqn{=} 
\sqrt{\frac{\Omega^2+6 \xi^2 h^2 / M_{\rm pl}^2}{\Omega^4}}\,.\label{Eq:chi}
\end{eqnarray}
One can calculate the slow-roll parameters as
\begin{eqnarray}
\epsilon \eqn{=} 
\frac{M_{\rm pl}^2}{2} \left( \frac{dU/d \chi}{U} \right)^2\,,
~~
\eta = M_{\rm pl}^2 \frac{d^2U/d \chi^2}{U}\,,
\end{eqnarray}
where $U(\chi)$ is obtained \footnote{In this analysis, we take only the quartic term of the Higgs field into account because the quadratic coupling can be negligible at the inflationary scale. Moreover, in our calculation, we use the $\beta$-functions of coupling constants including the self-coupling of the Higgs up to 2-loop level. This means that our analysis also contains quantum effects, which partially include effects from loop level potential.} as
\begin{eqnarray}
U(\chi) \eqn{\equiv} 
\frac{\lambda}{4\Omega(\chi)^4} \left( h (\chi)^2 - v_{\rm EW}^2 \right)^2\,.
\end{eqnarray}
Using slow-roll parameters, the spectral index and the tensor-to-scalar ratio 
are evaluated as $n_s = 1-6 \epsilon +2 \eta$ and $r=16 \epsilon$, respectively.
Finally, the number of e-foldings is given by 
\begin{eqnarray}
N \eqn{=} 
\int_{h_{\rm end}}^{h_0} \frac{1}{M_{\rm pl}^2} \frac{U}{dU/dh} \left( \frac{d\chi}{dh} \right)^2 dh\,,
\end{eqnarray}
where $h_0$ $(h_{\rm end})$ is the initial (final) value of $h$ corresponding to 
the beginning (end) of the inflation.
The end point $h_{\rm end}$ is defined as the slow-roll conditions $(\epsilon\,, |\eta| \ll 1)$ are broken.

The Higgs inflation can be realized even in the SM if the top mass is fine-tuned as $M_t = 171.079 \GeV$ for $m_H=125.6 \GeV$~\cite{Hamada:2013mya,Hamada:2014iga,Fairbairn:2014nxa}. 
With these values, the Higgs potential have a plateau and 
$r \simeq 0.2$ can be achieved by taking $\xi=7$~\cite{Hamada:2014iga}.
Even though this framework can accomplish a relevant amount of e-foldings, 
the required top mass, $M_t \simeq 171.1 \GeV$, is out of $M_t=173.34 \pm 0.76 \GeV$ \cite{ATLAS:2014wva}.
On the other hand, without the plateau, 
the enough amount of e-foldings can be obtained by assuming $\xi \sim \mathcal{O}(10^4)$.
However, this case is plagued with too tiny tensor-to-scalar ratio as order of $10^{-3}$ 
which is inconsistent with the result of BICEP2~\cite{Ade:2014xna}
On the other hand, the joint analysis of BICEP, Keck Array, and Planck concludes that this signal can be explained by the dust~\cite{Ade:2015tva}. However, this setup is still attractive in phenomenological points of view. Furthermore, we will show that the main results do not change so much even if we take the tensor-to-scalar ratio as $r=0.048$ which is the center value reported in Ref.~\cite{Ade:2015tva}.
Consequently, 
we extend the SM with a real scalar and right-handed neutrinos, 
in which an evolution of $\lambda$ (equivalent to the Higgs potential) is changed, 
in order to reproduce the values of cosmological parameters 
within the experimental range $M_t$.

%%%%%%%%%%%%%%%%%%%%%%%%%%%%%%%%%%
\section{Analyses}

In this section, we give results of our numerical analysis.
We solve RGEs\footnote{Here, we consider that this renormalization scale is the same as $h/\Omega$.} at 2-loop level for the $\beta$-functions of the relevant couplings in the model.\footnote{There are theoretical uncertainties between low and high energy parameters as discussed in \cite{Bezrukov:2010jz,Bezrukov:2014bra}. But we assume such uncertainties are enough small and can be neglected.}
As we discussed above, we analyze within the experimental ranges of the Higgs mass $m_h= 125.6 \pm 0.35 \GeV$ \cite{Hahn:2014qla} and the top mass $M_t=173.34 \pm 0.76 \GeV$ \cite{ATLAS:2014wva}. 
The detail of the 2-loop $\beta$-functions 

As we mentioned in the section II, 
we assume that the additional real scalar is DM.
In the case that the DM mass is greater than the Higgs mass ($m_S > m_h$), 
the portal coupling $k(M_Z)$ is well approximated as \cite{Cline:2013gha,Hamada:2014xka},
\begin{eqnarray}
\log_{10} k (M_Z) \simeq -3.63+1.04 \log_{10} \left( \frac{m_{\rm DM}} \GeV \right)\,,\label{Eq:DM_CONST}
\end{eqnarray}
where $m_{\rm DM}$ denotes the mass of DM given by $m_{\rm DM}^2 = m_S^2 + k v_{\rm EW}^2/2$.

\begin{figure}[t:]
\begin{center}
\includegraphics[width=6.5cm,clip,angle=-90]{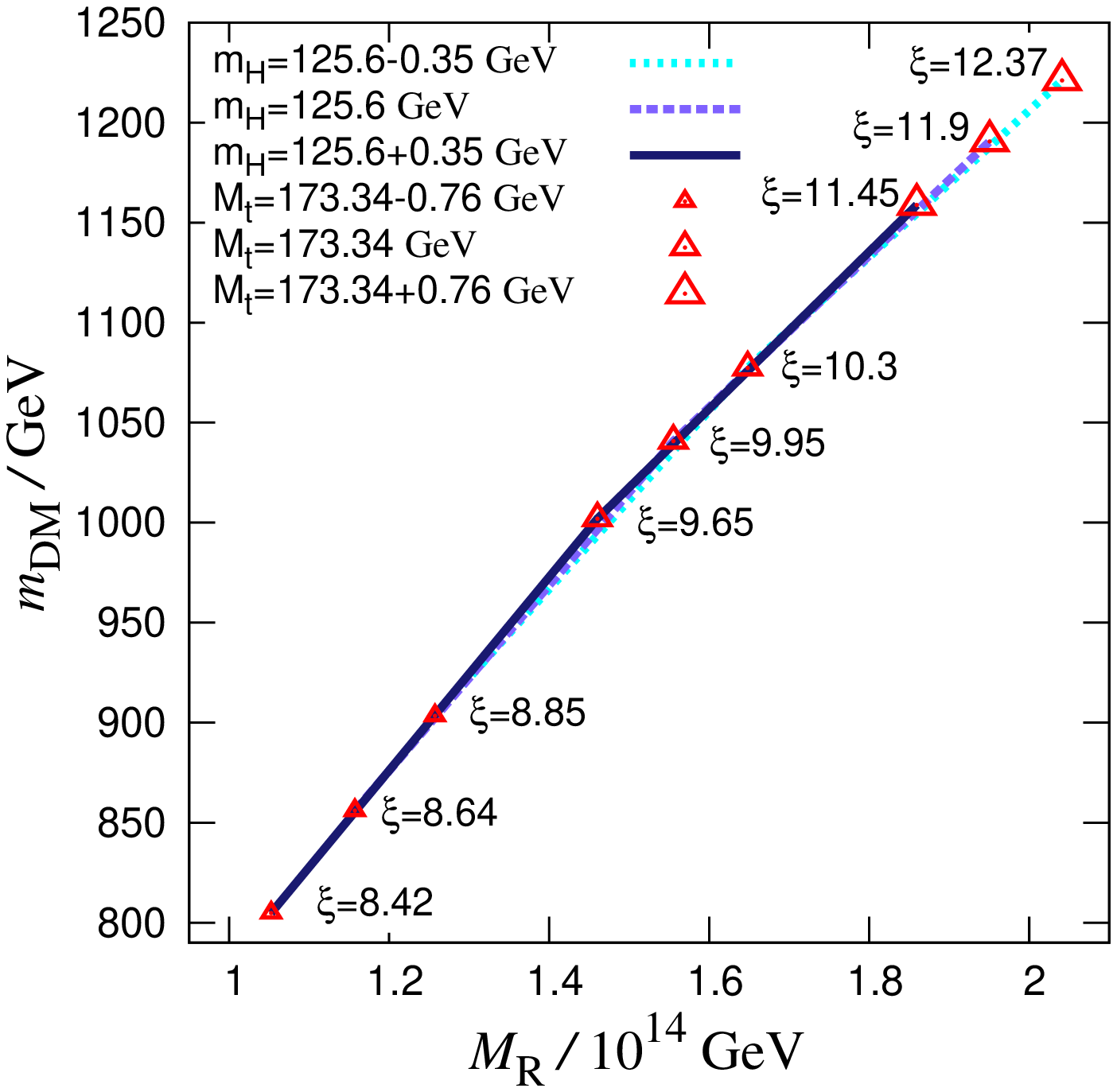}\hspace{-20mm}~~
\includegraphics[width=6.5cm,clip,angle=-90]{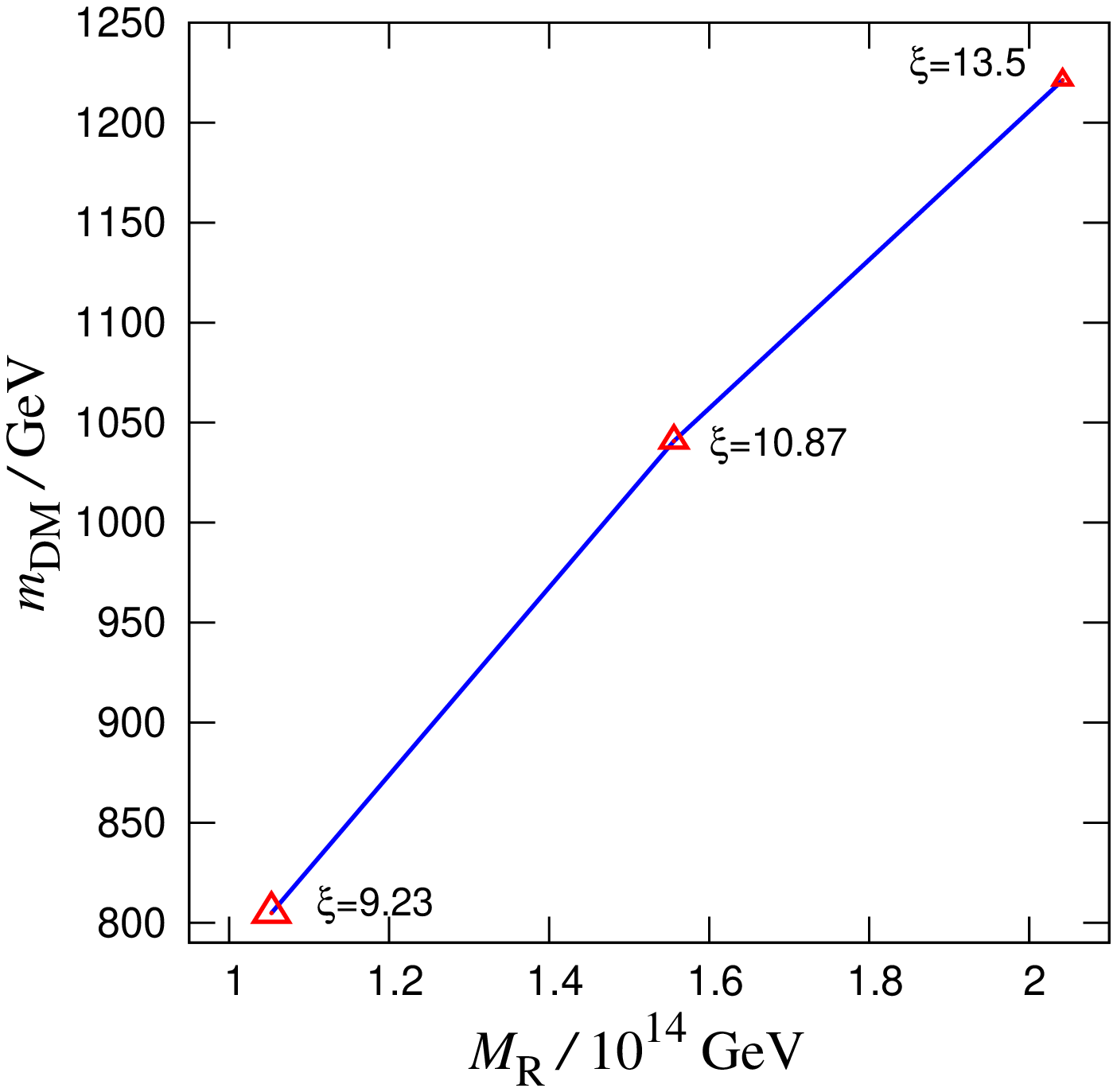}
\end{center}
\caption{$M_{\rm R}$-$m_{\rm DM}$ plane in this model, which reproduces the inflation parameters. {\bf Left:} The tensor-to-scalar ratio is fixed as $r=0.2$. {\bf Right:} The tensor-to-scalar ratio is fixed as $r=0.05$.}\label{Fig:MR_MS}
\end{figure}
Our results are shown in Fig.\ref{Fig:MR_MS}.
One can see that required values of $M_R$, $m_{\rm DM}$, and $\xi$ 
for reproducing suitable cosmological parameters are on a line in $M_R$-$m_{\rm DM}$ plane.
This is because the form of the Higgs potential is strictly constrained and it should be uniquely 
realized by taking suitable values of $M_R$, $m_{\rm DM}$, and $\xi$ for given $m_H$ and $M_t$.
A lighter (heavier) top (Higgs) mass gives lower bounds on $M_R$, $m_{\rm DM}$, and $\xi$ 
while a heavier (lighter) top (Higgs) mass leads to upper bounds on those parameters.

\section{Conclusions}
We have investigated the Higgs inflation model with the Higgs portal DM and the 
right-handed neutrinos by the use of the $\beta$-functions up to 2-loop level. In
 addition, the latest experimental errors of the top and Higgs masses 
 have been taken into account in the 
calculations. As a result, we pointed out that this inflation model can explain
 the results of cosmological observations $n_s = 0.9600 \pm 0.0071$, 
$r= 0.20^{+0.07}_{-0.06}$, and $52.3 \lesssim N \lesssim 59.7$ within regions of 
$805 \GeV \lesssim m_{\rm DM} \lesssim 1220 \GeV$ of the DM mass, 
$1.05 \times10^{14} \GeV \lesssim M_R \lesssim 2.04 \times10^{14} \GeV$ of 
the right-handed neutrino mass, and $8.42 \lesssim \xi \lesssim 12.4$ of the 
non-minimal Higgs coupling to the Ricci scalar with $m_H=125.6\pm0.35 \GeV$
 of the Higgs and $M_t=173.34\pm0.76 \GeV$ of the top masses. 
Furthermore, we have shown that the scale of right-handed neutrino and DM are not significantly changed even though the tensor-to-scalar ratio decreases as $r<0.12$.
On the other hand, the non-minimal coupling should be slightly larger than the $r=0.2$ case as $9.23 \lesssim \xi \lesssim 13.5$.
There is a strong correlation between the DM and the right-handed neutrino masses 
because the form of the Higgs potential is strictly constrained and it should be uniquely 
realized by taking suitable values of $M_R$, $m_{\rm DM}$, and $\xi$ for given $m_H$ and $M_t$.
The DM mass region in our analysis will be confirmed by the future DM detections.

%\section{Paper Submission}
%
%Authors should submit their papers to the ePrint arXiv 
%server.
%For instructions for submission, look at the help page of the arXiv.org.
%After successful submission to arXiv, the authors should
%inform the arXiv identifier (arXiv:xxxx.yyyy) to shindou@cc.kogakuin.ac.jp
%by e-mail with a subject as ``[HPNP15] arXiv:xxxx.yyyy''.
%\textcolor{red}{The deadline to do this is 30~April, 2015.}

% If you have acknowledgments, this puts in the proper section head.
%\bigskip % extra skip inserted
%%%%%%%%%%%%%%%%%%%%%%%%%%%%%%%%%%
\begin{acknowledgments}
The author would like to thank to all of organizers of HPNP2015 and great hospitality during the conference.
\end{acknowledgments}

\bigskip % extra skip inserted
% Create the reference section using BibTeX:
%\bibliography{basename of .bib file}

\end{document}